\documentclass{article} 
\usepackage{amsmath,amssymb,amsfonts}   
\usepackage{graphicx,bm}

\usepackage{amsmath,amssymb,amsfonts}   
\usepackage{graphicx,bm}

\newcommand{\calP}{{\mathcal P}}

\newcommand{\calF}{{\mathcal F}}

\newcommand{\calM}{{\mathcal M}}
\newcommand{\calL}{{\mathcal L}}
\newcommand{\calD}{{\mathcal D}}
\newcommand{\calW}{{\mathcal W}}
\newcommand{\R}{{\mathbb R}}
\renewcommand{\L}{{\mathbb L}}
\newcommand{\z}{\mathbf{z}}
\newcommand{\X}{\mathbf{X}}
\renewcommand{\P}{\mathbb{P}}
\newcommand{\PP}{\widetilde{P}}

\newcommand{\x}{\mathbf{x}}
\newcommand{\y}{\mathbf{y}}
\newcommand{\e}{{\mathrm e}}
\newcommand{\ellb}{{\bm \ell}}

\newcommand{\n}{\mathbf n}
\newcommand{\calT}{{\mathcal T}}
\renewcommand{\P}{\mathbb P}
\newcommand{\p}{\widetilde{p}}

\renewcommand{\S}{\widetilde{S}}
\newcommand{\ellh}{\widehat{\ell}}

\newcommand{\Markov}[2]{\underset{#1}{\overset{#2}{\rightleftharpoons}}}

\begin{document}

\title{Stochastically switching diffusion with partially reactive surfaces}
\author{ \em
P. C. Bressloff, \\ Department of Mathematics, 
University of Utah \\155 South 1400 East, Salt Lake City, UT 84112}

\maketitle
\begin{abstract}
In this paper we develop a hybrid version of the encounter-based approach to diffusion-mediated absorption at a reactive surface, which takes into account stochastic switching of a diffusing particle's conformational state. For simplicity, we consider a two-state model in which the probability of surface absorption depends on the current particle state and the amount of time the particle has spent in a neighborhood of the surface in each state. The latter is determined by a pair of local times $\ell_{n,t}$, $n=0,1$, which are Brownian functionals that keep track of particle-surface encounters over the time interval $[0,t]$.
We proceed by constructing a differential Chapman-Kolmogorov equation for a pair of generalized propagators $P_n(\x,\ell_0,\ell_1,t)$, where $P_n$ is the joint probability density for the set $(\X_t,\ell_{0,t},\ell_{1,t})$ when $N_t=n$, where $\X_t$ denotes the particle position and $N_t$ is the corresponding conformational state. Performing a double Laplace transform with respect to $\ell_0,\ell_1$ yields an effective system of equations describing diffusion in a bounded domain $\Omega$, in which there is switching between two Robin boundary conditions on $\partial \Omega$. The corresponding constant reactivities are $\kappa_j=D z_j$, $j=0,1$, where $z_j$ is the Laplace variable corresponding to $\ell_j$ and $D$ is the diffusivity. Given the solution for the propagators in Laplace space, we construct a corresponding probabilistic model for partial absorption, which requires finding the inverse Laplace transform with respect to $z_0,z_1$. We illustrate the theory by  considering diffusion of a particle on the half-line with the boundary at $x=0$ effectively switching between a totally reflecting and a partially absorbing state. We calculate the flux due to absorption and use this to compute the resulting MFPT in the presence of a renewal-based stochastic resetting protocol. The latter resets the position and conformational state of the particle as well as the corresponding local times. Finally, we indicate how to extend the analysis to higher spatial dimensions using the spectral theory of Dirichlet-to-Neumann operators. 
\end{abstract}

\section{Introduction}

An important quantity characterizing single-particle diffusion in a bounded domain $\Omega\subset \R^d$ is the first passage
time (FPT) for a particle to reach the boundary $\partial \Omega$ \cite{Metzler14,Benichou14}. Mathematically speaking, one can model the stochastic dynamics using standard Brownian motion, supplemented by the stopping condition that diffusion is terminated as soon as the particle reaches the boundary. The FPT is defined according to $\calT=\inf\{t>0, \X_t\in \partial \Omega\}$, where $\X_t$ is the position of the particle at time $t$. An alternative approach is to consider the probability density $p(\x,t)$ for particle position, which satisfies the diffusion equation in $\Omega$ with a Dirichlet (absorbing) boundary condition, namely, $p(\x,t)=0$ for all $\x\in \partial \Omega$.

One limitation of the above picture is that it ignores what happens after the particle reaches the boundary surface. In many applications, the surface acts as a reactive boundary layer, within which the particle can bind, undergo a change in conformational state, participate in a chemical reaction, be transported to the exterior of the domain through a membrane pore, or be destroyed. The particle could represent a protein within a cell, a bacterium searching for some resource within a confinement domain, or a chemical reactant interacting with a catalytic substrate \cite{Bressloff13,Grebenkov19a,Bressloff22}. Irrespective of the details, a typical surface reaction is unlikely to be instantaneous, but require an alternating sequence of periods of bulk diffusion interspersed with local surface interactions before the final ``absorption'' event is realized. In other words, the boundary $\partial \Omega$ acts as a partially absorbing surface \cite{Note1}.

The simplest mathematical implementation of partial absorption is to replace the Dirichlet boundary condition in the diffusion equation by the Robin boundary condition $D\nabla p(\x,t)\cdot \n+\kappa_0 p(\x,t)=0$ for all $\x\in \partial \Omega$. Here $D$ is the diffusivity, $\kappa_0$ is a constant reactivity that characterizes the rate at which absorption occurs, and $\n$ is the outward unit normal at a point on the boundary. The Dirichlet boundary condition is recovered in the limit $\kappa_0\rightarrow \infty$, whereas the boundary becomes totally reflecting when $\kappa_0=0$. However, in order to implement the Robin boundary condition at the level of single-particle trajectories, it is necessary to modify the underlying stochastic differential equation. For example, the effects of a totally reflecting boundary can be incorporated by considering so-called reflected Brownian motion. This involves the introduction of a Brownian functional known as the boundary local time $\ell_t$, which characterizes the amount of time that a Brownian particle spends in the neighborhood of points on the boundary \cite{Levy39,McKean75,Majumdar05}. Heuristically speaking, the differential of the local time generates an impulsive kick whenever the particle encounters the boundary, leading to the so-called stochastic Skorokhod equation \cite{Freidlin85}. It is also possible to construct a probabilistic implementation of the Robin boundary condition for partially reflected Brownian motion \cite{Papanicolaou90,Milshtein95} and more general continuous stochastic processes \cite{Singer08}.

The assumption that surface absorption can be modeled in terms of a constant reactivity $\kappa_0$ is itself an idealization of more realistic surface-based reactions \cite{Bartholomew01,Filoche08}. 
For example, the surface may need to be progressively
activated by repeated encounters with a diffusing
particle. Alternatively, an initially highly reactive surface may become less active due to multiple interactions with the particle (passivation).
Both cases can be modeled by taking the reactivity to be a function of the boundary local time.
Recently, a probabilistic framework for analyzing this more general class of partially absorbing boundary has been developed using a so-called encounter-based approach \cite{Grebenkov20,Grebenkov22}. The underlying idea is that the Robin boundary condition is equivalent to imposing a stopping condition for the local time $\ell_t$ of the particle:  ${\mathcal T}=\inf\{t>0:\ \ell_t >\widehat{\ell}\}$,
where $\widehat{\ell}$ is a stopping local time with an exponential probability distribution. That is, $\P[\ellh>\ell]\equiv \Psi(\ell)=\e^{-\gamma \ell}$ with $\gamma =\kappa_0/D$. The corresponding probability density can then be written in the form $ p(\x,t)=\int_0^{\infty} \Psi(\ell)P(\x,\ell,t)d\ell$, where $P(\x,\ell,t)$ is the joint probability density or generalized propagator for the pair $(\X_t,\ell_t)$ in the case of a perfectly reflecting boundary. The crucial observation is that the propagator $P$ satisfies a boundary value problem (BVP) that is independent of the details of the surface reactions. (The propagator BVP can be derived using integral representations \cite{Grebenkov20} or the Feynman-Kac formula \cite{Bressloff22a}.) Hence, a much more general class of surface reactions can be incorporated by considering appropriately defined non-exponential distributions $\Psi(\ell)$. For example, in the case of a reactivity $\kappa(\ell)$ that depends on the local time, we have $\Psi(\ell)=\exp\left (-D^{-1}\int_0^{\ell}\kappa(\ell')d\ell'\right )$.

Another source of complexity in diffusion mediated surface reactions is stochastic switching. A classical example is the membrane transport of charged particles via voltage-gated or ligand-gated ion channels that randomly switch between open and closed states \cite{Reingruber10,Bressloff15a,Bressloff15c}. Each channel effectively acts as a semi-permeable local boundary, which is absorbing (reflecting) whenever the channel is open (closed). Moreover, the random switching could be due to intrinsic properties of the channels or due to changes in the conformational state of the diffusing molecules. The two scenarios are statistically equivalent at the single-particle level. (On the other hand, for a population of independently diffusing particles, there are additional correlations in the case of switching gates due to the fact that all particles experience the same switching environment \cite{Bressloff15a}.) Irrespective of the mechanism, randomly switching boundary conditions can be modeled in terms of a stochastic hybrid system involving a set of probability densities $p_j(\x,t)$, $j=1,\ldots,N$, where $N$ is the number of discrete states. The probability densities evolve according to a differential Chapman-Kolmogorov (CK) equation that couples diffusion with a Markov chain that takes into account transitions between the states. Such transitions could occur during bulk diffusion or be induced by surface-particle interactions.

In this paper we develop a hybrid version of the encounter-based approach to partially absorbing surfaces that takes into account stochastic switching of the diffusing particle's conformational state. For simplicity, we consider a two-state model in which the probability of absorption at the boundary depends on the current particle state $N_t\in \{0,1\}$ and the amount of time the particle has spent in a neighborhood of the boundary in each state. In addition, we assume that transitions between the conformational states only occur when the particle is diffusing in the bulk domain. We begin by briefly describing the BVP for the generalized propagator without switching and showing how to incorporate a probabilistic rule for partial absorption (Sect. II), following along the lines of Ref. \cite{Grebenkov20}. We then generalize the theory to the case of switching boundary conditions, at least one of which is partially absorbing (Sect. III). First, we introduce a pair of local times $\ell_{j,t}$, $j=0,1$, that keep track of the time spent in a neighborhood of the boundary $\partial \Omega$ while in state $j$. Second, we define a CK equation for a pair of generalized propagators $P_j(\x,\ell_0,\ell_1,t)$, where $P_j$ is the joint probability density for the set $(\X_t,\ell_{0,t},\ell_{1,t})$ when $N_t=j$. Performing a double Laplace transform with respect to $\ell_0,\ell_1$ yields a CK equation describing diffusion in a bounded domain in which there is switching between two Robin boundary conditions on $\partial \Omega$ with constant reactivities $\kappa_j=D z_j$, $j=0,1$, where $z_j$ is the Laplace variable corresponding to $\ell_j$. Third, given the solution of the propagator BVP in Laplace space, we construct the corresponding probabilistic model for partial absorption, which requires finding the inverse Laplace transform with respect to $z_0,z_1$. 

Next, we define various quantities of interest such as the surface flux and the mean first passage time (MFPT) for absorption when $\Omega$ is bounded (Sect. IV).  We also consider the complementary problem in which the particle diffuses in the unbounded domain exterior to $\Omega$, that is, $\Omega^c=\R^d\backslash \Omega$. In this case, the MFPT to be absorbed by $\partial \Omega$ is infinite. One mechanism for obtaining a finite MFPT is to reset the particle state at a random sequence of times, which is typically taken to be a Poisson process with rate $r$ (see the review \cite{Evans20}). We assume that the corresponding local times also reset, which ensures that resetting is governed by a renewal process. This then allows us to calculate the MFPT in terms of the surface flux without resetting. 

We illustrate the theory by considering diffusion of a particle on the half-line with the boundary at $x=0$ effectively switching between a totally reflecting and a partially absorbing state (Sect. V). We solve the associated one-dimensional (1D) BVP for the propagators in Laplace space, invert with respect to the Laplace variables $z_j$, and then determine the effective flux due to absorption. The flux is then used to compute the MFPT for absorption in the presence of stochastic resetting. In particular, we explore how the MFPT depends on various model parameters, including the resetting rate $r$, the switching rates, and the surface reactivities. Finally, we indicate how to extend the analysis to higher spatial dimensions using the spectral decomposition of a pair of Dirichlet-to-Neumann operators (Sect. VI). This generalizes the analysis previously developed for non-switching systems \cite{Grebenkov20}.

 \section{Generalized propagator BVP without switching}
Consider a particle diffusing inside a bounded domain $\Omega\subset \R^d$ with a totally reflecting boundary $\partial \Omega$, see left-hand  panel of Fig. \ref{fig1}.  
 Let $\X_t$ denote the position of the particle at time $t$ and denote the boundary local time by $\ell_t$. The latter is defined according to
\begin{equation}
\label{loc}
\ell_t=\lim_{h\rightarrow 0} \frac{D}{h} \int_0^tH(h-\mbox{dist}(\X_{\tau},\partial \Omega))d\tau,
\end{equation}
where $H$ is the Heaviside function. Note that $\ell_t$, which has units of length due to the additional factor of $D$, specifies the amount of time that the particle spends in an infinitesimal neighborhood of the surface $\partial \Omega$. Eq. (\ref{loc}) implies that $\ell_t$ is a non-decreasing stochastic process, which remains at zero until the first encounter with the boundary. Although each surface encounter takes place over an infinitely short time interval, the particle returns to the surface multiple times before reentering the bulk, so that there is a measurable change in $\ell_t$. 
 It can be shown that the propagator satisfies a BVP of the form \cite{Grebenkov20,Bressloff22a}
\begin{subequations}
\label{Ploc}
\begin{align}
 &\frac{\partial P(\x,\ell,t|\x_0)}{\partial t}=D\nabla^2 P(\x,\ell,t|\x_0),\ \x \in \Omega,\\
 &-D\nabla P(\x,\ell,t|\x_0) \cdot \n= D P(\x,\ell=0,t|\x_0) \ \delta(\ell)   +D\frac{\partial}{\partial \ell} P(\x,\ell,t|\x_0),  \x\in \partial \Omega.
\end{align}
The unit normal $\n$ on $\partial \Omega$ is directed towards the exterior of $\Omega$.
These equations are supplemented by the ``nitial conditions'' $P(\x,\ell,0|\x_0)=\delta(\x-\x_0)\delta(\ell)$ and
\begin{equation}
P(\x,\ell=0,t|\x_0)=-\nabla p_{\infty}(\x,t|\x_0)\cdot \n \mbox{ for } \x\in \partial \Omega, 
\end{equation}
\end{subequations}
where $p_{\infty}$ is the probability density in the case of a totally absorbing surface $\partial \Omega$:
\begin{subequations} 
\label{pinf}
\begin{align}
 	&\frac{\partial p_{\infty}(\x,t|\x_0)}{\partial t} = D\nabla^2 p_{\infty}(\x,t|\x_0), \, \x\in \Omega ,\\
	 &p_{\infty}(\x,t|\x_0)=0,\  \x\in \partial \Omega,\ p_{\infty}(\x,0|\x_0)=\delta(\x-\x_0).
	\end{align}
	\end{subequations} 
An intuitive interpretation of the boundary condition (\ref{Ploc}b) is that the rate at which the local time increases is proportional to the flux into the boundary when $\ell_t >0$. However, this process only starts once the particle has reached the surface for the first time, which is identical to the case of a totally absorbing surface.

The construction of the marginal probability density $p(\x,t|\x_0)$ in the case of a partially absorbing surface proceeds as follows \cite{Grebenkov20,Grebenkov22}.  Introducing the double Laplace transform 
\begin{equation}
 \label{dLT}
 \calP(\x,z,s|\x_0)\equiv \int_0^{\infty}\e^{-z\ell}\int_0^{\infty}\e^{-st}P(\x,\ell,t|\x_0)dtd\ell,
 \end{equation} 
we have
\begin{subequations}
\label{PlocLT}
\begin{eqnarray}
 &&D\nabla^2 \calP(\x,z,s|\x_0)-s\calP(\x,z,s|\x_0)=-\delta(\x-\x_0),\ \x \in \Omega,\nonumber \\ & \\
&&-\nabla \calP(\x,z,s|\x_0) \cdot \n=z\calP(\x,z,s|\x_0) ,\ \x\in \partial \Omega .
\end{eqnarray}
\end{subequations}
If $z\equiv \gamma_0=\kappa_0/D$ for some constant $\kappa_0$ then
the BVP (\ref{PlocLT}) is identical to the $s$-Laplace transformed diffusion equation in the case of a Robin boundary condition on $\partial \Omega$ with a constant rate of reactivity $\kappa_0$. In other words, the solution of the classical BVP
\begin{subequations}
\label{Robin}
\begin{align}
 &\frac{\partial p(\x,t|\x_0)}{\partial t}=D\nabla^2 p(\x,t|\x_0),\ \x \in \Omega,\\
 &-D\nabla p(\x,t|\x_0) \cdot \n= \kappa_0 p(\x,t|\x_0),\ \x\in \partial \Omega,
\end{align}
\end{subequations}
can be expressed as
\begin{equation}
\label{bob}
p(\x,t|\x_0)=\int_0^{\infty} \e^{-\gamma_0\ell}P(\x,\ell,t|\x_0)d\ell =\PP(\x,\gamma_0,t|\x_0).
\end{equation}
This, in turn, is equivalent to introducing
an absorption stopping time,
\begin{equation}
\label{TA}
{\mathcal T}=\inf\{t>0:\ \ell_t >\widehat{\ell}\},
\end{equation}
 with $\widehat{\ell}$ an exponentially distributed random variable that represents a stopping local time \cite{Grebenkov20}. That is, $\Psi(\ell)\equiv \P[\widehat{\ell}>\ell]=\e^{-\gamma_0\ell}$.   
The advantage of formulating the Robin boundary condition in terms of the generalized propagator is that one can consider a more general probability distribution $\Psi(\ell) $ for the stopping local time $\ellh$ such that \cite{Grebenkov19a,Grebenkov20,Grebenkov22}
  \begin{equation}
  \label{oo}
  p (\x,t|\x_0)=\int_0^{\infty} \Psi(\ell)P(\x,\ell,t|\x_0)d\ell \ \mbox{ for } \x \in \Omega .
  \end{equation}
This accommodates a much wider class of surface reactions where, for example, the reactivity $\kappa(\ell)$ depends on the local time $\ell$ (or the number of surface encounters):
\begin{equation}
\label{kaell}
\Psi(\ell)=\exp\left (-\frac{1}{D}\int_0^{\ell}\kappa(\ell')d\ell'\right ).
\end{equation}
 Laplace transforming equation (\ref{oo}) with respect to $t$ gives
  \begin{equation}
  \label{oo2}
  \p  (\x,s|\x_0)=\int_0^{\infty} \Psi(\ell){\mathcal L}_{\ell}^{-1}[\calP(\x,z,s|\x_0)] d\ell \ \mbox{ for } \x \in \Omega ,
  \end{equation}
  where $\calP(\x,z,s|\x_0)$ is the solution of the Robin BVP given by equations (\ref{PlocLT}). That is, the marginal density $\p(\x,s|\x_0)$ for a general distribution $\Psi(\ell)$ can be obtained by solving a classical Robin BVP with effective reactivity $\kappa=zD$ and then inverting the Laplace transform with respect to $z$.  
  
  \setcounter{equation}{0}
   \section{Generalized propagator BVP with switching}
   
   \begin{figure*}
 \centering
\includegraphics[width=12cm]{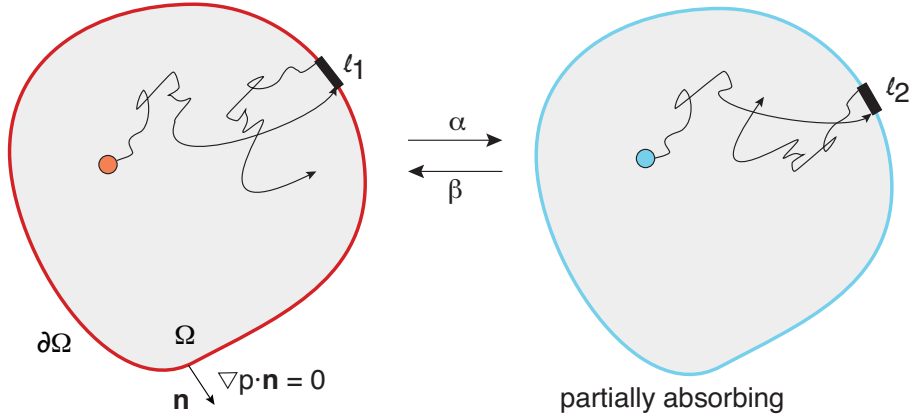}
\caption{Schematic diagram of a Brownian particle diffusing in a bounded domain $\Omega$ and switching between two conformational states, such that the boundary $\partial \Omega$ is either reflecting ($N(t)=0$) or partially absorbing boundary ($N(t)=1$). Here $N(t)$ is a two-state Markov chain with transition rates $\alpha,\beta$. Prior to absorption, each encounter between the particle in state $N(t)=$ and the boundary increases the corresponding local time $\ell_{j,t}$. 
} \label{fig1}
\end{figure*}

Now suppose that the particle switches between two conformational states labeled by the discrete random variable $N_t \in \{0,1\}$. Moreover, we assume that the probability of absorption at the boundary depends on the current particle state and the amount of time the particle has spent in a neighborhood of the boundary in each state. In Fig. \ref{fig1} we show the example of switching between
a totally reflecting state ($N_t=0)$ and a partially absorbing state ($N_t=1$), although one could consider both states to be partially absorbing. The state variable $N_t$ evolves according to a two-state Markov chain,
\[0\Markov{\beta}{\alpha} 1,\]
with constant transition rates $\alpha,\beta$.
Introduce the pair of boundary local times
\begin{equation}
\label{locpair}
\ell_{j,t}=\lim_{h\rightarrow 0} \frac{D}{h} \int_0^tH(h-\mbox{dist}(\X_{\tau},\partial \Omega))\delta_{N_{\tau},j}d\tau .
\end{equation}
That is, $\ell_{j,t}$ is the local time accumulated over the interval $[0,t]$ when the boundary is in the state $n\in \{0,1\}$. We also set ${\bm \ell}_t=(\ell_{0,t},\ell_{1,t})$. Introduce a corresponding pair of propagators
\begin{align}
&P_j(\x,\ellb,t)d\x\, d\ellb =\P[\x<X_t<\x+d\x,\ellb<\ellb_t<\ellb+d\ellb,\, N_t=j] .\nonumber 
\end{align}
For notational convenience, we drop the explicit dependence on the initial conditions
\begin{equation}
\X_0=\x_0,\quad \ellb_0=0,\quad \P[N_0=j]=\rho_j,
\end{equation}
where $\rho_j$, $j=0,1$, is the stationary distribution of the Markov chain:
\begin{equation}
\rho_0=\frac{\beta}{\alpha+\beta},\quad \rho_1=\frac{\alpha}{\alpha+\beta}.
\end{equation}

The pair of propagators satisfy the system of differential CK equations
\begin{subequations}
\label{swPloc}
\begin{align}
 \frac{\partial P_0(\x,{\bm \ell},t)}{\partial t}&=D\nabla^2 P_0(\x,{\bm \ell},t)-\alpha P_0(\x,{\bm \ell},t )\ +\beta P_1(\x,{\bm \ell},t) , \, \x\in \Omega,\,\\
 \frac{\partial P_1(\x,{\bm \ell},t)}{\partial t}&=D\nabla^2 P_1(\x,{\bm \ell},t)+\alpha P_0(\x,{\bm \ell},t )-\beta P_1(\x,{\bm \ell},t ) , \, \x\in \Omega,\\ 
-D\nabla P_j(\x,{\bm \ell},t) \cdot \n&= D P_j(\x,\ellb,t) \ \delta(\ell_j) +D\frac{\partial}{\partial \ell_j} P_j(\x,\ellb,t) \,  ,\x\in \partial \Omega   \end{align}
\end{subequations}
for $j=0,1$.
The corresponding initial conditions are
\begin{equation}
P_j(\x,{\bm \ell},0 )=\rho_j\delta(\x-\x_0)\delta(\ellb),\ \x \in \Omega,\, j=0,1.
\end{equation}
Finally, we introduce the marginal propagator
\begin{align}
&P(\x,\ellb,t)d\x\, d\ellb  =\P[\x<X_t<\x+d\x,\ellb<\ellb_t<\ellb+d\ellb,] \nonumber 
\end{align}
such that
\begin{equation}
\label{Pmar}
P(\x,\ellb,t)=P_0(\x,\ellb,t)+P_1(\x,\ellb,t).
\end{equation}
For simplicity, we assume that the diffusivity $D$ is the same in both conformational states. 

Introducing the triple Laplace transform 
\begin{equation}
 \label{swdLT}
 \calP_j(\x,\z,s)\equiv \int_0^{\infty}\e^{-\z\cdot \ellb}\int_0^{\infty}\e^{-st}P_j(\x,\ell,t)dtd\ellb,
 \end{equation} 
we have
\begin{subequations}
\label{swPlocLT}
\begin{eqnarray}
 &&D\nabla^2 \calP_0(\x,\z,s)-(s+\alpha)\calP_0(\x,\z,s)+ \beta \calP_1(\x,\z,s) =-\rho_0\delta(\x-\x_0),\ \x \in \Omega,\nonumber \\ \\
 &&D\nabla^2 \calP_1(\x,\z,s)+\alpha\calP_0(\x,\z,s)-(s+ \beta) \calP_1(\x,\z,s) =-\rho_1\delta(\x-\x_0),\ \x \in \Omega, \nonumber \\ \\
&&-\nabla \calP_j(\x,\z,s) \cdot \n=z_j\calP_j(\x,\z,s) ,\ \x\in \partial \Omega 
\end{eqnarray}
\end{subequations}
for $j=0,1$.
For fixed $z_j$, the BVP (\ref{swPlocLT}) is precisely the CK equation for a particle diffusing in a bounded domain where there is switching between two Robin boundary conditions on $\partial \Omega$ with constant reactivities $\kappa_j=D z_j$, $j=0,1$. In particular, for the switching system shown in Fig. \ref{fig1} we would have $z_0=0$ and $z_1>0$, and the corresponding pair of marginal densities would be
\begin{align}
p_j(\x,t)&=\int_0^{\infty}d\ell_0 \int_0^{\infty}d\ell_1\, \e^{-z_1\ell_1} P_j(\x,\ellb,t)=\calP_j(\x,0,z_1,t).
\end{align} 

In order to generalize the switching Robin boundary conditions, we assume that absorption occurs as soon as either local time crosses its own independent threshold:
\begin{equation}
\label{TA2}
{\mathcal T}=\inf\{t>0:\ \{\ell_{0,t} >\widehat{\ell}_0\}\lor \{\ell_{1,t} >\widehat{\ell}_1\}  \},
\end{equation}
where $\widehat{\ell}_j$ is an independent random variable with probability density $\psi_j(\ell)$.
Since the local times are nondecreasing, it follows that the condition $t < {\mathcal T}$ is equivalent to the condition 
$\ell_{j,t}  <\widehat{\ell}_j$ for $j=0,1$.
This implies that 
\begin{align*}
p(\x,t)d\x&=\P[\X_t \in (\x,\x+d\x), \ \{\ell_{0,t }< \widehat{\ell}_0\}\land \{\ell_{1,t }< \widehat{\ell}_1\}]\\
&=\int_0^{\infty} du \, \psi_0(u)\int_0^{\infty} du' \, \psi_1(u')\P[\X_t \in (\x,\x+d\x), \ \ell_{0,t} <u,\, \ell_{1,t} < u ' ].
\end{align*}
That is,
\begin{align}
p(\x,t)&=\int_0^{\infty} du \, \psi_0(u)\int_0^{\infty} du' \, \psi_1(u')\int_{0}^{u}d\ell\int_{0}^{u'}d\ell'P(\x,\ell,\ell',t),
\end{align}
where $P(\x,\ellb,t)$ is the marginal propagator (\ref{Pmar}). Using the identity
\begin{equation}
\label{id}
\int_0^{\infty}du\ f(u)\int_0^{u} d\ell \ g(\ell)=\int_0^{\infty}d\ell \ g(\ell) \int_{\ell}^{\infty} du \ f(u)
\end{equation}
for arbitrary integrable functions $f,g$, we have
\begin{align}
\label{peep}
p(\x,t)&=\int_0^{\infty} du \, \psi_0(u)\int_{0}^{u} d\ell \int_0^{\infty} d\ell'P(\x,\ell,\ell',t) \, \int_{\ell'}^{\infty}du'\, \psi_1(u')\nonumber \\
&=\int_0^{\infty} du \, \psi_0(u)\int_{0}^{u} d\ell \int_0^{\infty}d\ell'  \Psi_1(\ell')P(\x,\ell,\ell',t) \nonumber \\
&=\int_0^{\infty}d\ell \int_0^{\infty}d\ell'  \Psi_1(\ell')P(\x,\ell,\ell',t) \, \int_{\ell}^{\infty}du \, \psi_0(u) \nonumber \\
& =\int_0^{\infty}d\ell \int_0^{\infty}d\ell'  \Psi_0(\ell)\Psi_1(\ell')P(\x,\ell,\ell',t),
\end{align}
where $\Psi_j(\ell)=\int_{\ell}^{\infty}\psi_j(u)du$.
 Eq. (\ref{peep}) can be rewritten in the more compact form
\begin{equation}
\label{peepg}
p(\x,t)= \int \calD \ellb \,\Psi(\ellb) P(\x,\ellb,t),\quad \calD \ellb  \equiv\int_0^{\infty}d\ell_0 \int_0^{\infty}d\ell_1 
\end{equation}
with $\Psi(\ellb)=\Psi_0(\ell_0)\Psi_1(\ell_1)$. Laplace transforming with respect to $t$ shows that
\begin{equation}
\p(\x,s)=\int \calD \ellb \,   \Psi(\ellb) \, \calL^{-1}_{\ell_0} \calL^{-1}_{\ell_1}[\calP(\x,\z,s)],
\end{equation}
where $\calP(\x,\z,s)$ is the solution to the hybrid BVP (\ref{swPlocLT}), and $\calL^{-1}$ indicates the inverse Laplace transform operator. 

\begin{figure*}[t!]
 \centering
\includegraphics[width=12cm]{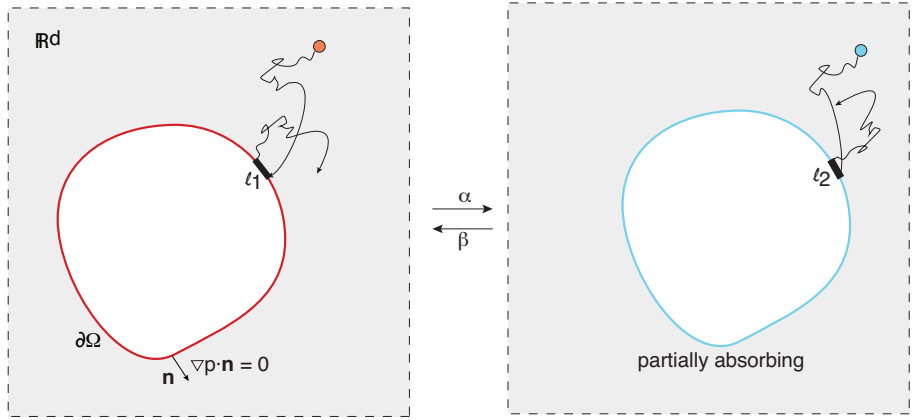}
\caption{Same as Fig. \ref{fig1} except that the particle diffuses in the unbounded domain exterior to $\Omega$. } \label{fig2}
\end{figure*}

\setcounter{equation}{0}

\section{Survival probability and first-passage time (FPT)  density}

Let $S(\x_0,t)$ denote the survival probability that the particle hasn't been absorbed in the time interval $[0,t]$,
\begin{equation}
S(\x_0,t)=\int_{\Omega} p(\x,t)d\x.
\end{equation}
Differentiating both sides with respect to $t$ and using Eq. (\ref{peepg}) gives
\begin{align}
\label{St}
\frac{\partial S(\x_0,t)}{\partial t}&=\int_{\Omega} \left \{\int \calD \ellb \, \Psi(\ellb) \frac{\partial P(\x,\ellb,t)}{\partial t}\right \}d\x.
\end{align}
Adding Eqs. (\ref{swPloc}a) and (\ref{swPloc}b) shows that the marginal propagator $ P(\x,\ellb,t)$ satisfies
\begin{align}
 \frac{\partial P(\x,{\bm \ell},t)}{\partial t}&=D\nabla^2 P(\x,{\bm \ell},t) .
 \end{align}
 Assuming that we can reverse the order of integration in Eq. (\ref{St}),
 \begin{align}
&\frac{\partial S(\x_0,t)}{\partial t}= \int \calD \ellb \, \Psi(\ellb) \int_{\Omega}\nabla^2 P(\x,\ellb,t) d\x=\int \calD \ellb \,  \Psi(\ellb) \int_{\partial \Omega}\nabla P(\x,\ellb,t) \cdot \n d\x\nonumber \\
&=- \int \calD \ellb \,   \Psi(\ellb) \int_{\partial \Omega}d\x\sum_{j=0,1}  \bigg (P_j(\x,\ellb,t) \ \delta(\ell_j)+\frac{\partial}{\partial \ell_j} P_j(\x,\ellb,t) \bigg ) \nonumber \\
&=\int_{\partial \Omega}d\x\int \calD \ellb \,   \left [ \frac{\partial \Psi(\ellb)}{\partial \ell_0}P_0(\x,\ellb,t)+\frac{\partial \Psi(\ellb)}{\partial \ell_1}P_1(\x,\ellb,t)\right ]\nonumber \\
&\equiv -J_0(\x_0,t)-J_1(\x_0,t),
\label{SJ}
\end{align}
where $J_j(\x_0,t)$ is the probability flux due to absorption in state $n$:
\begin{equation}
J_j(\x_0,t)=-\int_{\partial \Omega}d\x\int \calD \ellb \,   \frac{\partial \Psi(\ellb)}{\partial \ell_j}P_j(\x,\ellb,t).
\end{equation}
In the case of the product rule $\Psi(ellb)=\Psi_0(\ell_0)\Psi_1(\ell_1)$, we have
\begin{equation}
\frac{\partial \calF(\ellb)}{\partial \ell_j}=-\psi_j(\ell_j)\Psi_{1-j}(\ell_{1-j}),\ j=0,1.
\end{equation}
The total probability flux is $J(\x_0,t)=J_0(\x_0,t)+J_1(\x_0,t)$.

Laplace transforming equation (\ref{SJ}) with respect to $t$ and noting that $S(\x_0,0)=1$ gives
\begin{equation}
\label{QL}
s\widetilde{S}(\x_0,s)-1=- \widetilde{J}(\x_0,s).
\end{equation}
Since  $-\partial S/\partial t$ is the probability density of the stopping time $\calT$, equation (\ref{TA}), we see that the MFPT (if it exists) is
\begin{align}
T(\x_0)&=-\int_0^{\infty}t\frac{\partial S(\x_0,t)}{\partial t}dt =\int_0^{\infty} S(\x_0,t)dt \nonumber \\
&=\S(\x_0,0)=-\left .\frac{\partial \widetilde{J}(\x_0,s)}{\partial s}\right |_{s=0}.
\label{MFPT1}
\end{align}
Similarly, higher order moments of the FPT density are given by higher order derivatives of $\widetilde{J}(\x_0,s)$. We conclude that the statistics of absorption 
can be determined from the Laplace transformed fluxes
\begin{align}
\label{JJ}
\widetilde{J}_j(\x_0,s)&=\int_{\partial \Omega}d\x\int \calD \ellb\,  \frac{\partial\Psi(\ellb)}{\partial \ell_j}\, \calL^{-1}_{\ell_0} \calL^{-1}_{\ell_1}[\calP_j(\x,\z,s)].
\end{align}
In turn, the latter is computed by solving the propagator BVP (\ref{swPlocLT}) for $\calP_0,\calP_1$ and then inverting the Laplace transforms with respect to $z_0,z_1$. 

So far we have assumed that the particle diffuses within the bounded domain $\Omega$. A complementary scenario is shown in Fig. 2, where the particle now diffuses in the unbounded domain exterior to $\Omega$. The only modification of the propagator BVPs (\ref{swPloc}) and (\ref{swPlocLT}) is that $\Omega$ is replaced by $\Omega^c= \R^d\backslash \Omega$, since $\partial \Omega^c=\partial \Omega$. However, it is well known that the MFPT for diffusion in an unbounded domain is infinite. One way to obtain a finite MFPT is to introduce some form of stochastic resetting (see the recent review \cite{Evans20}). In the case of diffusion with resetting in $\R^d$, one typically assumes that the position of the particle is instantaneously reset to its initial position $\x_0$, say, at a random sequence of times generated by a Poisson process with rate $r$ \cite{Evans11a,Evans11b,Evans14}. We have previously shown how to modify the resetting rule in the case of a boundary that randomly switches between a totally absorbing state and a totally reflecting state \cite{Bressloff20}, see also \cite{Boyer21}. More recently, we have also considered diffusion with resetting in a domain with a partially absorbing boundary and no switching \cite{Bressloff22c}. Based on these studies, suppose that prior to absorption, the following resetting protocol occurs at a Poisson rate $r$ \cite{note2}:
\begin{equation}
\X_t\rightarrow \x_0,\quad \ellb_t\rightarrow (0,0),\quad N_t \rightarrow j \mbox{ with probability } \rho_j.
\end{equation}
Using renewal theory, one finds that the Laplace transform of the survival probability with resetting, which we denote by $S_r(\x_0,t)$, is related to the corresponding function without resetting according to \cite{Bressloff22c}
\begin{equation}
\S_r(\x_0,s)=\frac{\S(\x_0,r+s)}{1-r\S(\x_0,r+s)}.
\end{equation}
Taking the limit $s\rightarrow 0$ and denoting the corresponding MFPT with resetting by $T_r(\x_0)$, we have
\begin{equation}
\label{Tr}
T_r(\x_0)=\frac{\S(\x_0,r)}{1-r\S(\x_0,r)}=\frac{1-\widetilde{J}(x_0,r)}{r\widetilde{J}(x_0,r)}.
\end{equation}
Therefore, Eq. (\ref{JJ}) can also be used to calculate $T_r(\x_0)$.

\setcounter{equation}{0}
\section{Diffusion on the half-line} 

We now illustrate the basic theory developed in the previous sections by considering diffusion in the semi-finite interval $\Omega =[0,\infty)$ with the boundary $\partial \Omega =\{0\}$ effectively switching between a totally reflecting and a partially absorbing state. This is a 1D version of the scenario shown in Fig. \ref{fig2}. We first solve the hybrid propagator BVP and then invert with respect to $\z$ in order to determine the flux through $x=0$ using Eq. (\ref{JJ}). This will then be used to calculate the MFPT with resetting according to Eq. (\ref{Tr}). Even for this relatively simple geometry, the analysis is quite involved.

\subsection{Calculation of the propagators}

The 1D version of the Laplace transformed BVP (\ref{swPlocLT}) takes the form
\begin{subequations}
\label{1D}
\begin{eqnarray}
 &&D\frac{\partial^2 \calP_0(x,\z,s)}{\partial x^2}-(s+\alpha)\calP_0(x,\z,s)+ \beta \calP_1(x,\z,s)\nonumber \\
 &&\quad =-\rho_0\delta(x-x_0),\ 0<x<\infty, \\
 &&D\frac{\partial^2 \calP_1(x,\z,s)}{\partial x^2}+\alpha \calP_0(x,\z,s)-(s+ \beta) \calP_1(x,\z,s)\nonumber \\
 &&\quad =-\rho_1\delta(x-x_0),\ 0<x<\infty, \\
&&\left .\frac{\partial \calP_j(x,\z,s)}{\partial x}\right |_{x=0}=z_j\calP_j(0,\z,s),\quad j=0,1 .
\end{eqnarray}
\end{subequations} 
Set 
\begin{equation}
\calP_j(x,\z,s)=\rho_jG(x,s|\x_0)+\calF_j(x,\z,s),
\end{equation}
where
$G$ is the modified Helmholtz Green's function with
\begin{eqnarray}
\label{Grr}
 D\frac{\partial^2 G(x,s|x_0)}{\partial x^2}-sG(x,s|x_0)=-\delta(x-x_0)
\end{eqnarray}
for $0<x<\infty$ and $G(0,s|x_0)=0$. It is straightforward to show from the method of images that
\begin{align}
G(x,s|x_0)=\frac{1}{2\sqrt{sD}}\left [\e^{-\mu(s)|x-x_0|}-\e^{-\mu(s) (x+x_0)}\right ]\\
\nonumber
\end{align}
with $\mu(s)=\sqrt{s/D}$.
Given the definition of $G$, it follows that $\calF_j(x,\z,s)$, $j=0,1$, satisfy the system of equations
and
\begin{subequations}
\label{1DF}
\begin{align}
& D\frac{\partial^2 \calF_0(x,\z,s)}{\partial x^2}-(s+\alpha)\calF_0(x,\z,s)+ \beta \calF_1(x,\z,s)=0,\,\\
 &D\frac{\partial^2 \calF_1(x,\z,s)}{\partial x^2}+\alpha \calF_0(x,\z,s)-(s+ \beta) \calF_1(x,\z,s)=0,\\
&\left .\frac{\partial \calF_j(x,\z,s)}{\partial x}\right |_{x=0}-z_j\calF_j(0,\z,s)=-\rho_j\left .\frac{\partial G(x,s|x_0)}{\partial x}\right |_{x=0},\,  \x\in \partial \Omega,\ j=0,1.
\end{align}
\end{subequations}

Adding Eqs. (\ref{1DF}a) and (\ref{1DF}b) implies that
\begin{eqnarray}
\label{Pw}
 &&D\frac{\partial^2 \calF(x,\z,s)}{\partial x^2}-s\calF(x,\z,s)=-\delta(x-x_0)
\end{eqnarray}
for $0<x<\infty$ and $\calF=\calF_0+\calF_1$. However, we do not have an explicit boundary condition for $\calF$. Therefore, we impose the inhomogeneous Dirichlet boundary condition $\calF(0,\z,s)=f(\z,s)$ with $f$ to be determined.
The equation for $\calF$ can then be solved using Green's second identity,
\begin{align}
\label{calPsol}
\calF(x,\z,s)&=Df(\z,s)\partial_{y}G(y,s|x)|_{y=0}=f(\z,s)\e^{-\mu(s)x}.
\end{align}
 The next step is to set $\calF_0=\calF-\calF_1$ in Eq. (\ref{1DF}b):
\begin{eqnarray}
 &&D\frac{\partial^2 \calF_1(x,\z,s)}{\partial x^2}-(s+\alpha+\beta)\calF_1(x,\z,s) =-\alpha \calF(x,\z,s) .\nonumber \\
 \label{Pw0}
\end{eqnarray} 
Imposing the Dirichlet boundary condition $\calF_1(0,\z,s)=f_1(\z,s)$ for a second unknown function $f_1$, we obtain the solution
\begin{align}
\label{Pww}
 \calF_1(x,\z,s)&=f_1(\z,s)e^{-\nu(s)x}\\
 &\quad +\alpha\int_0^{\infty} G(x,s+\alpha+\beta|y)\calF(y,\z,s)dy \nonumber,
 \end{align}
with $\nu(s)=\sqrt{[s+\alpha+\beta]/D}$. Substituting for $\calF$ using Eq. (\ref{calPsol}) gives
\begin{align}
\calF_1(x,\z,s)&=f_1(\z,s)\e^{-\nu(s)x}+ {K}(x,s)f(\z,s),\label{calPsol0}
\end{align}
 with
 \begin{align}
{K}(x,s)&=\alpha\int_0^{\infty} G(x,s+\alpha+\beta|y)e^{-\mu(s)y}dy.
 \end{align}

 The final step is to determine the unknown functions $f_1(\z,s)$ and $f(\z,s)$ by imposing the pair of boundary conditions (\ref{1DF}c). First consider the case $j=1$. Since $K(0,s)=0$,
it follows from Eq. (\ref{calPsol0}) that $\calF_1(0,\z,s)=f_1(\z,s)$ and thus 
\begin{align*}
z_1 f_1(\z,s)&=\partial_x\calF_1(0,\z,s)+\rho_1 \partial_{x}G(0,s|x_0) \nonumber \\
&= -\nu(s)f_1(\z,s)+ \frac{\rho_1}{D}\e^{-\mu(s)x_0}+{K}(s)f(\z,s),
\end{align*}
where $'$ indicates differentiation with respect to $x$ and $K(s)\equiv K'(0,s)$ with
 \begin{align}
 \label{Kp}
{K}(s)&= \frac{\alpha}{D} \int_0^{\infty}\e^{-\nu(s)y}\e^{-\mu(s)y}dy
= \frac{1}{D}\frac{\alpha}{\mu(s)+\nu(s)}.
\end{align}
  We thus obtain the first condition relating $f$ and $f_1$:
 \begin{equation}
 \label{cond1}
 [z_1+\nu(s)]f_1(\z,s)=  \frac{\rho_1}{D}\e^{-\mu(s)x_0}+K(s)f(\z,s).
 \end{equation}
 The second condition is obtained by setting $j=0$ and $\calF_0=\calF-\calF_1$ in Eq. (\ref{1DF}c):
 \begin{align*}
 \partial_x\calF(0,\z,s)-\partial_x\calF_1(0,\z,s)&=z_0[\calF(0,\z,s)-\calF_1(0,\z,s)]-\rho_0 \partial_{x}G(0,s|x_0) ,
 \end{align*}
 which can be rearranged to give
 \begin{equation}
\partial_x\calF(0,\z,s)-z_0\calF(0,\z,s)=(z_1-z_0)f_1(\z,s)-\frac{1}{D}\e^{-\mu(s)x_0}.
\end{equation} 
Substituting for $\calF$ using Eq. (\ref{calPsol}) then gives
\begin{align}
\label{cond2}
\frac{1}{D}\e^{-\mu(s)x_0}-[\mu(s)+z_0)]f(\z,s)=(z_1-z_0)f_1(\z,s).
\end{align}
Finally, combining equations (\ref{cond1}) and (\ref{cond2}) yields the solutions
\begin{equation}
\label{f}
f(\z,s)=\frac{D^{-1}[\nu(s)+z_0]\e^{-\mu(s)x_0}}
{[\nu(s)+z_1][\mu(s)+z_0]+K(s)(z_1-z_0)},
\end{equation}
and
\begin{align}
f_1(\z,s)&=\frac{ \e^{-\mu(s)x_0}}{D}\left [\nu(s)+z_1-\frac{(z_0-z_1)K(s)}{\mu(s)+z_0}\right ]^{-1}\left  [\rho_1+ \frac{K(s)}{\mu(s)+z_0 }\right ].
\label{f1}
\end{align}

\subsection{Calculation of the absorption flux and the MFPT with resetting}
 
 Since the boundary at $x=0$ is totally reflecting when $N_t=0$, it follows that the stopping local time distribution $\Psi_0(\ell)=1$ for all $\ell$ and $J_0(x_0,t)=0$.
 Hence, the total flux due to absorption is
 \begin{equation}
J(x_0,t)= \int \calD \ellb \,  \psi_1(\ell_1) P_1(0,\ellb,t),
\end{equation}
where $\psi_1(\ell)=-\Psi'_1(\ell)$ is the stopping local time density for the absorbing state. Laplace transforming with respect to time $t$, we have
  \begin{equation}
  \label{flux}
\widetilde{J}(x_0,s)= \int_0^{\infty}d\ell_1 \,  \psi_1(\ell_1) \calL_{\ell_1}^{-1}\calP_1(0,z_0=0,z_1,s).
\end{equation}
Substituting for $\calP_1$ using Eq. (\ref{calPsol0}) gives
\begin{align}
\widetilde{J}(x_0,s)&= \int_0^{\infty}d\ell_1 \,  \psi_1(\ell_1)\calL_{\ell_1}^{-1} f_1(z_0=0,z_1,s) .
  \end{align}
  Setting $z_0=0$ in Eq. (\ref{f1}) yields
   \begin{align}
   \label{f1z0}
  f_1(z_0=0,z_1,s)=\frac{ \overline{\Theta}(s)}{z_1+\overline{\nu}(s)},
   \end{align}
   where
   \begin{equation}
   \overline{\Theta}(s)=\left (1+\frac{\rho_1\mu(s)}{K(s)}\right )\frac{\e^{-\mu(s)x_0}}{D},
   \end{equation}
   and
   \begin{equation}
   \overline{\nu}(s)=\frac{\nu(s)\mu(s)}{\mu(s)+K(s)},\quad \nu(s)=\sqrt{\frac{s+\alpha+\beta}{D}}.
   \end{equation}
  Hence,
   \begin{align}
  \widetilde{J}(x_0,s)&= \int_0^{\infty}d\ell \,  \psi_1(\ell) \overline{\Theta}(s)\e^{-\overline{\nu}(s)\ell} =\widetilde{\psi}_1(\overline{\nu}(s))\overline{\Theta}(s) .
     \label{Jthet}
  \end{align}

We can now investigate the behavior of the MFPT with resetting $T_r$ by substituting Eq. (\ref{Jthet}) into Eq. (\ref{Tr}) for $s=r$, where $r$ is the resetting rate. For the sake of illustration, we take $\psi_1$ to be the gamma distribution:
\begin{equation}
\label{psigam}
\psi_{\rm gam}(\ell)=\frac{\gamma(\gamma \ell)^{a-1}\e^{-\gamma \ell}}{\Gamma(a)},\quad \widetilde{\psi}_{\rm gam}(z)=\left (\frac{\gamma}{\gamma+z}\right )^{a},\ a >0,
\end{equation}
where $\Gamma(a)$ is the gamma function  
\begin{equation}
\Gamma(a)=\int_0^{\infty}\e^{-t}t^{a-1}dt.
\end{equation}
The parameter $\gamma$ determines the effective absorption rate so that the surface $\partial \Omega$ is totally reflecting in the limit $\gamma\rightarrow 0$ and totally absorbing in the limit $\gamma \rightarrow \infty$ when $N_t=1$. (In the latter case, if $x_0>0$ then the particle is absorbed as soon as it reaches $x=0$.) If $a=1$ then $\psi_{\rm gam}$ reduces to the exponential distribution with constant reactivity $\gamma$, that is, $\psi_{\rm gam}(\ell)|_{a =1}=\gamma \e^{-\gamma \ell}$. The parameter $a$ thus characterizes the deviation of $\psi_{\rm gam}(\ell)$ from the exponential case. If $a <1$ ($a>1$) then $\psi_{\rm gam}(\ell)$ decreases more rapidly (slowly) as a function of the local time $\ell$.

Clearly $T_r$ is going to be an increasing function of $a$ and a decreasing function of $\gamma$. It will also decrease when the relative amount of time that the boundary is in the partially absorbing state ($\rho_1$) increases. Since the domain $\Omega$ is unbounded, we also expect the MFPT $T_r$ to be a unimodal function of the resetting rate $r$ with a minimum at some optimal rate $r_{\rm opt}$. What is less clear is how $r_{\rm opt}$ varies with other model parameters. We also want to explore how the MFPT depends on the relative rate of switching for fixed $\rho_1$, which is determined by $\Gamma=\alpha+\beta$, and to compare the results for a partially absorbing state (finite $\gamma)$ with a totally absorbing state ($\gamma \rightarrow \infty$). In order to perform the latter comparison, we introduce the normalized MFPT 
\begin{equation}
\Delta{T}_r(\x_0) \equiv \frac{T_r(\x_0)}{T_{r,\infty}(\x_0)},\quad T_{r,\infty}(\x_0)=\lim_{\gamma \rightarrow \infty}T_r(\x_0),
\end{equation}
where  $T_{r,\infty}(\x_0)$ is the MFPT in the case of switching between a totally reflecting and a totally absorbing boundary condition. Since $T_{r,\infty}(\x_0)$ is independent of the parameters $(a,\gamma)$, this essentially allows us to separate out the dependence on the gamma distribution. Moreover, although $T_r$ and $T_{r,\infty}$ blow up in the limit $\rho_1\rightarrow 0$ (no absorption), we find that their ratio converges to a finite value. Therefore, we set
\begin{equation}
\overline{T}_r(\x_0)=\frac{\Delta {T}_r(\x_0)}{\lim_{\rho_1\rightarrow 0} \Delta {T}_r(\x_0)} .
\end{equation}
Finally, we fix the length and time scales by setting $D=1$ and $x_0=1$.

\begin{figure}[t!]
  \includegraphics[width=12cm]{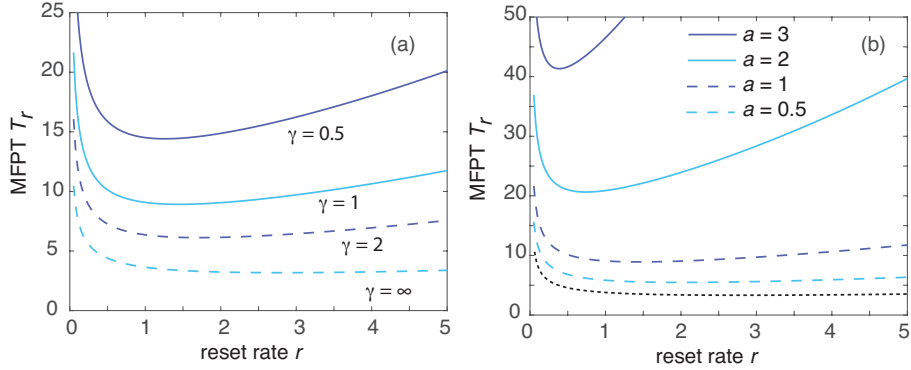}
  \caption{MFPT $T_r$ for diffusion in the half-line with the boundary condition at $x=0$ switching between a totally reflecting state and a partially absorbing state governed by the gamma distribution (\ref{psigam}) with parameters $(a,\gamma)$. We take the switching rates $\alpha=\beta=0.5$ and set $D=1$, $x_0=1$. (a) Plot of $T_r$ as a function of $r$ for various values of $\gamma$ and $a=1$, which corresponds to the exponential distribution (constant reactivity). (b) Corresponding plots $T_r$ for various values of $a$ and $\gamma=1$. The dotted curve corresponds to the totally absorbing case ($\gamma \rightarrow \infty$).}
  \label{fig3}
\end{figure}

\begin{figure}[t!]
  \includegraphics[width=12cm]{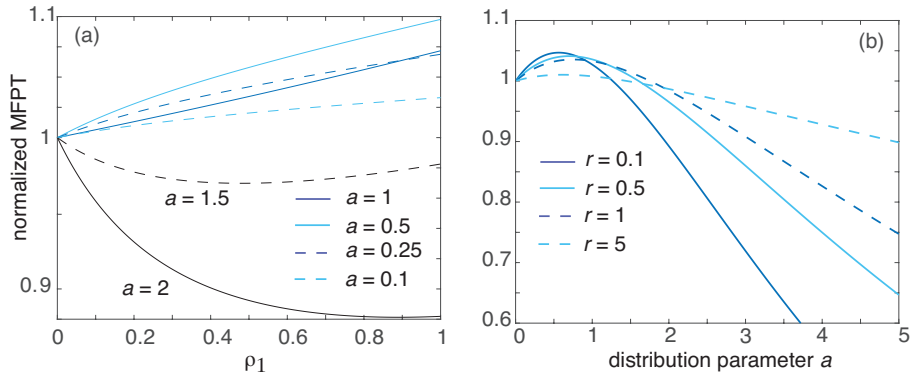}
  \caption{Normalized MFPT $\overline{T}_r$ for diffusion in the half-line with the boundary condition at $x=0$ switching between a totally reflecting state and a partially absorbing state governed by the gamma distribution (\ref{psigam}) with parameters $(a,\gamma)$. We take $\alpha+\beta=1$, $D=1$, $\gamma=1$ and $x_0=1$. (a) Plot of $\overline{T}_r$ as a function of $\rho_1=\alpha/(\alpha+\beta)$ for various values of $a$ and $r=1$. (b) Corresponding plots of $\overline{T}_r$ as a function of the distribution parameter $a$ for various values of $r$ with $\rho_1=0.5$.}
  \label{fig4}
\end{figure}

 \begin{figure}[t!]
  \includegraphics[width=12cm]{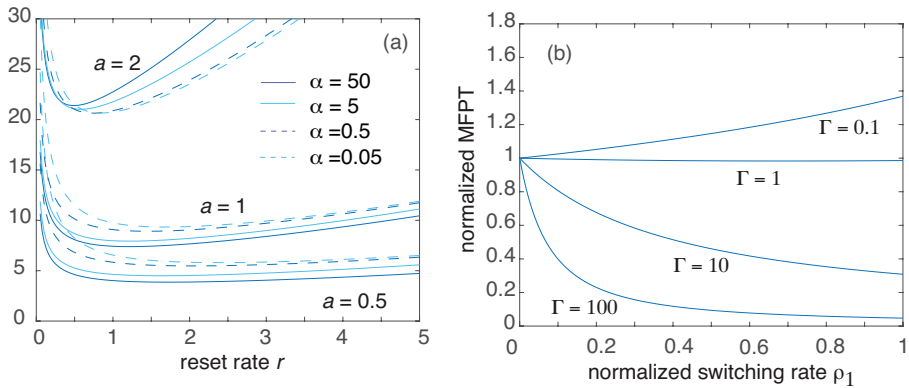}
  \caption{(a) Plot of MFPT $T_r$ as a function of the resetting rate $r$ for various values of $a$ and $\alpha$ with $\rho_1=0.5$. (b) Plot of normalized MFPT $\overline{T}_r$ as a function of $\rho_1$ for various values of $\Gamma=\alpha+\beta$ with $r=1$ and $a=2$. Other parameters are $D=1$, $\gamma=1$ and $x_0=1$. }
  \label{fig5}
\end{figure}

In Fig \ref{fig3} we plot $T_r$ as a function of the resetting rate for various combinations of $(a,\gamma)$ and $\alpha=\beta=0.5$.
It can be seen that $T_r$ is indeed a unimodal function of $r$ with a minimum at an optimal rate $r_{\rm opt}$. Moreover, we find that  $r_{\rm opt}$ is an increasing function of $\gamma$ and a decreasing function of $a$. The curves converge in the limit $\gamma \rightarrow \infty$ for fixed $a$. In Fig. \ref{fig4}(a) we plot the normalized MFPT $\overline{T}_r$ as a function of $\rho_1$ for different values of the parameter $a$. It can be seen that up to a critical value of $a$ (which depends on $r$ and $\gamma$), increasing $\rho_1$ increases the normalized MFPT. This implies that the MFPT $T_{r,\infty}$ decreases more quickly than $T_r$ as the relative time spent in the state $n=1$ increases. Beyond this critical value of $a$, the normalized MFPT is a decreasing function of $\rho_1$. It also follows that the normalized MFPT $\overline{T}_r$ is a non-monotonic function of $a$ for fixed $\rho_1$ as illustrated in Fig. \ref{fig4}(b). In Fig. \ref{fig5}(a) we plot $T_r$ as a function of $r$ for different values of $a$ and $\Gamma$. We also take $\rho_1=0.5$ so that $\alpha = 0.5 \Gamma$. We observe a nontrivial crossover effect, namely, increasing the switching rate $\Gamma$ decreases $T_r$ for small $a$ but increases $T_r$ for large $a$. Finally, in Fig. \ref{fig5}(b) we plot the normalized MFPT $\overline{T}_r$ as a function of $\rho_1$ for various $\Gamma$, showing a switch in behavior as $\Gamma$ increases. This is analogous to the switch in behavior in Fig. \ref{fig4}(a).

 \subsection{Fast switching limit}
 
One subtle feature of switching systems is what happens in the fast switching limit $\alpha,\beta\rightarrow \infty$. In order to investigate such a limit we introduce the scalings $\alpha,\beta\rightarrow \alpha/\epsilon,\beta/\epsilon$ with $\alpha,\beta =O(1)$.
The Laplace transformed BVP (\ref{swPlocLT}) for diffusion in $\Omega \subset \R^d$ becomes
\begin{subequations}
\label{1Dfast}
\begin{eqnarray}
 &&D\frac{\partial^2 \calP_0(x,\z,s)}{\partial x^2}-\left [s+\frac{\alpha}{\epsilon}\right ]\calP_0(x,\z,s)+\frac{\beta}{\epsilon} \calP_1(x,\z,s)\nonumber \\
 &&\quad =-\rho_0\delta(x-x_0),\ 0<x<\infty, \\
 &&D\frac{\partial^2 \calP_1(x,\z,s)}{\partial x^2}+\frac{\alpha}{\epsilon} \calP_0(x,\z,s)-\left [s+\frac{\beta}{\epsilon}\right ] \calP_1(x,\z,s)\nonumber \\
 &&\quad =-\rho_1\delta(x-x_0),\ 0<x<\infty, \\
&&\left .\frac{\partial \calP_j(x,\z,s)}{\partial x}\right |_{x=0}=z_j\calP_j(0,\z,s),\quad j=0,1 .
\end{eqnarray}
\end{subequations} 
It is tempting to carry out an adiabatic approximation of Eqs. (\ref{1Dfast}a,b) by decomposing the density $\calP_j$ as
\begin{equation}
\label{deco}
\calP_j(\x,\z,s)=\rho_j\calP(\x,\z,s) +\epsilon \calW_j(\x,\z,s),
\end{equation}
where $\sum_{j=0,1}  \calW_j=0$ and $\calP=\calP_0+\calP_1$ with
\begin{align}
\label{BBC}
D\nabla^2\calP(\x,\z,s) -s\calP(\x,\z,s)=-\delta(\x-\x_0).
\end{align}
The problem is that the leading order approximation $\calP_j=\rho_j\calP$ does not satisfy the pair of boundary conditions (\ref{1Dfast}c) when $z_0\neq z_1$. However, it is possible to satisfy the single boundary condition that is obtained by summing with respect to $j$:
\begin{equation}
\left .\frac{\partial \calP(x,\z,s)}{\partial x}\right |_{x=0}=\sum_{j=0,1}z_j\calP_j(0,\z,s).
\end{equation}
Setting $\calP_j=\rho_j\calP$ gives
\begin{equation}
\label{BBC2}
\left .\frac{\partial \calP(x,\z,s)}{\partial x}\right |_{x=0}=\overline{z} \calP(0,\z,s),\quad \overline{z}=\sum_{j=0,1}\rho_j z_j.
\end{equation}
We thus have a closed equation for $\calP$ given by Eqs. (\ref{BBC}) and (\ref{BBC2}). Finally, the solution for the individual components $\calP_n$ in the fast switching limit can be obtained using matched asymptotics. That is, the outer solution $\calP_j(x,\z,s)=\rho_j \calP(x,\z,s)$ for $x>0$ is matched with an inner solution that holds within a boundary layer around $x=0$ so that it satisfies the remaining boundary condition. Analogous methods have previously been applied to models of molecular motor transport \cite{Newby10,Zmurchok17} and Brownian motion with switching diffusivities \cite{Bressloff19a}. The need for a boundary layer also arises when deriving a Robin boundary condition via the temporal homogenization of a stochastically switching boundary \cite{Lawley15}.

Rather than implementing the matched asymptotic analysis here, we apply the fast switching limit directly to the solutions (\ref{f}) and (\ref{f1}). First, taking the limits $\alpha,\beta\rightarrow \infty$ with $\rho_0,\rho_1$ fixed in (\ref{f}) gives
 \begin{align}
f(\z,s)&\sim \frac{1}{D} \frac{ \nu(s) \e^{-\mu(s)x_0}}
{\nu(s)[\mu(s)+z_0]+\alpha(z_1-z_0)/\nu(s)D}\nonumber \\
&\sim \frac{1}{D}\frac{  \e^{-\mu(s)x_0}}
{ \mu(s)+\overline{z}}.
\label{fcond1}
\end{align}
We thus obtain the following adiabatic approximation \cite{note3}
\begin{align}
\label{2calPsol}
\calP(x,\z,s)&\sim G(x,s|x_0)+\frac{1}{D}\frac{  \e^{-\mu(s)x_0}}
{ \mu(s)+\overline{z}}\e^{-\mu(s)x},
\end{align}
which is precisely the solution to Eqs. (\ref{BBC}) and (\ref{BBC2}). Similarly, taking the fast switching limit of Eq. (\ref{f1}) shows that
\begin{align}
\calP_1(0,\z,s)\equiv f_1(\z,s)&\sim \frac{\rho_1}{D}\frac{  \e^{-\mu(s)x_0}}
{ \mu(s)+\overline{z}}.
\label{fcond2}
\end{align}
That is, $\calP_1(0,\z,s)=\rho_1 \calP(0,\z,s)$. (On the other hand, $\partial_x\calP_1(0,\z,s)\neq \rho_1 \partial_x\calP(0,\z,s)$, which reflects the existence of a boundary layer of size $1/\sqrt{\epsilon}$ that ensures the correct boundary condition for $\calP_1$ is satisfied.)

Given the approximation (\ref{fcond2}), the associated flux in Eq. (\ref{flux}) becomes
 \begin{align}
\widetilde{J}(x_0,s)&= \frac{\rho_1\e^{-\mu(s)x_0}}{D}\int_0^{\infty}d\ell_1 \,  \psi_1(\ell_1) \calL_{\ell_1}^{-1}\frac{1}{\mu(s)+\rho_1 z_1}\nonumber \\
&= \frac{\e^{-\mu(s)x_0}}{D}\widetilde{\psi}_1(\mu(s)/\rho_1).
\label{flux2}
\end{align}
It can be checked numerically that the solution (\ref{Jthet}) converges to the solution (\ref{flux2}) in the fast switching limit. We conclude that the only difference between the flux into a partially absorbing surface without switching and the corresponding flux due to fast switching between a totally reflecting surface and a partially absorbing surface is the scaling $\widetilde{\psi}_1(\mu(s))\rightarrow \widetilde{\psi}_1(\mu(s)/\rho_1)$, where $\widetilde{\psi}_1$ is the Laplace transform of the stopping local time density and $\rho_1$ is the relative amount of time that the boundary is partially absorbing. 

\setcounter{equation}{0}
\section{Spectral theory in higher spatial dimensions}

It turns out the the analysis of the 1D BVP (\ref{1D}) developed in Sect. V can be extended to higher-dimensions using spectral theory. It has previously been shown that in the absence of switching, one can use the spectral decomposition of a so-called Dirichlet-to-Neumann operator defined on the boundary $\partial \Omega$ \cite{Grebenkov20}. The basic idea is to decompose the solution of the propagator BVP (\ref{PlocLT}) according to
\begin{equation}
\calP(\x,z,s|\x_0)=G(\x,s|\x_0)+\calF(\x,z,s|\x_0),
\end{equation}
where $G$ is the higher-dimensional version of the modified Helmholtz Green's function (\ref{Grr}) and
\begin{subequations}
\label{homT}
\begin{align}
 &D\nabla^2 \calF(\x,z,s|\x_0)-s\calF(\x,z,s|\x_0)=0,\ \x \in \Omega,\\
&\nabla \calF(\x,z,s|\x_0) \cdot \n+z\calF(\x,z,s|\x_0)\nonumber \\
& \quad =- \nabla G(\x,s|\x_0) \cdot \n,\   \x\in \partial \Omega.
\end{align}
\end{subequations} 
Replacing the Robin boundary condition by the Dirichlet condition $\calF(\x,z,s|\x_0)=f(\x,z,s)$ leads to the equation
\begin{equation}
\label{fL}
\L_s[f](\x,z,s)+zf(\x,z,s)=-\partial_{\sigma}G(\x,s|\x_0),
\end{equation}
where $\L_s$ is the Dirichlet-to-Neumann operator
\begin{align}
\label{DtoN}
\L_s[f](\x,s) &=-D\partial_{\sigma}\int_{\partial \Omega}\partial_{\sigma'}G(\x',s|\x)f(\x',s)d\x',
\end{align}
$\partial_{\sigma}\equiv \n \cdot \nabla_{\x}$ and $\partial_{\sigma'}\equiv \n \cdot \nabla_{\x'}$.
When the surface $\partial \Omega$ is bounded, the Dirichlet-to-Neumann operator $\L_s$ has a discrete spectrum. That is, there exist countable set of eigenvalues $\lambda_n(s)$ and eigenfunctions $v_n(\x,s)$ satisfying (for fixed $s$)
\begin{equation}
\label{eig}
\L_s v_n(\x,s)=\lambda_n(s)v_n(\x,s),\quad n\geq 0.
\end{equation}
(It can be shown that the eigenvalues are non-negative and that the eigenfunctions form a complete orthonormal basis in $L_2(\partial \Omega)$. Hence, we can solve equation (\ref{fL}) by introducing an eigenfunction expansion 
\begin{equation}
\label{eig2}
f(\x,z,s)=\sum_{m=0}^{\infty}f_m(z,s) v_m(\x,s).
\end{equation}
 This yields the result \cite{Grebenkov20}
\begin{equation}
 \label{spec2}
\calP(\x,z,s|\x_0)=G(\x,s|\x_0)+\frac{1}{D}\sum_{n=0}^{\infty} \frac{{\mathcal V}_n(\x,s){\mathcal V}^*_n(\x_0,s)}{\lambda_n(s)+z},
\end{equation}
where
 \begin{align}
 \label{calV}
{\mathcal V}_n(\x,s)&=-D\int_{\partial \Omega} v_n(\x',s)\partial_{\sigma'}G(\x',s|\x)d\x'.
\end{align}  

An analogous spectral decomposition can be performed in the case of the solution of the switching system (\ref{swPlocLT}) by following the same sequence of steps as the 1D case. 
\medskip

\noindent (i) Set (after dropping the explicit dependence on initial conditions)
\begin{equation}
\calP_j(\x,\z,s)=\rho_jG(\x,s|\x_0)+\calF_j(\x,\z,s),
\end{equation}
where
\begin{subequations}
\label{swhomT}
\begin{align}
 &D\nabla^2 \calF_0(\x,z,s)-(s+\alpha)\calF_0(\x,z,s)+\beta \calF_1(\x,z,s)=0,\\
  &D\nabla^2 \calF_1(\x,z,s)+\alpha\calF_0(\x,z,s)-(s+\beta )\calF_1(\x,z,s)=0,\\
&\nabla \calF_j(\x,z,s) \cdot \n+z_j\calF_j(\x,z,s)=-\rho_j \nabla G(\x,s|\x_0) \cdot \n,\   \x\in \partial \Omega.
\end{align}
\end{subequations} 
 Adding Eqs. (\ref{swhomT}a,b) and setting $\calF=\calF_0+\calF_1$ gives
 \begin{equation}
D\nabla^2 \calF(\x,z,s)-s\calF(\x,z,s)=0,\ \x \in \Omega,
\end{equation}
which is supplemented by the Dirichlet boundary condition $\calF(\x,z,s)=f(\x,\z,s)$ for $\x\in \partial \Omega$ and an unknown function $f$.
We thus obtain the formal solution
\begin{equation}
\label{swF}
\calF(\x,z,s)=-D\int_{\partial \Omega} \partial_{\sigma'} G(\x',s|\x)f(\x',\z,s)d\x'.
\end{equation}

\noindent (ii) Set $\calF_0=\calF-\calF_1$ in Eq.  (\ref{swhomT}b) so that
\begin{align}
D\nabla^2 \calF_1(\x,z,s)-(s+\alpha+\beta )\calF_1(\x,z,s)= -\alpha\calF(\x,z,s)
\end{align}
for $\x\in \Omega$, and impose the second Dirichlet boundary condition $\calF_1=f_1$ for $\x\in  \partial \Omega$. This leads to the formal solution
\begin{align}
\label{swF1}
\calF_1(\x,z,s)&=-D\int_{\partial \Omega} \partial_{\sigma'} G(\x',s+\alpha+\beta|\x)f_1(\x',\z,s)d\x'\nonumber \\  
&\quad +\alpha\int_{\Omega} G(\x,s+\alpha+\beta|\y)\calF(\y,z,s)d\y \ \mbox{for } \x\in \Omega. 
\end{align}

\noindent (iii) Derive a pair of self-consistency conditions for the unknown functions $f$ and $f_1$ by imposing the Robin boundary conditions (\ref{swhomT}c). First, substituting Eq. (\ref{swF1}) into (\ref{swhomT}c) with $j=1$, we have
\begin{align}
\label{woo}
&z_1f_1(\x,\z,s)+\alpha \int_{\Omega} \partial_{\sigma} G(\x,s+\alpha+\beta|\y)\calF(\y,z,s)d\y \\
&\ - D\partial_\sigma \int_{\partial \Omega} \partial_{\sigma'} G(\x',s+\alpha+\beta|\x)f_1(\x',\z,s)d\x'\ =-\rho_j \partial_{\sigma} G(\x,s|\x_0),\ \x \in \partial \Omega.\nonumber
\end{align}
Denote the integral on the first line by ${\mathcal I}$. Substituting for $\calF$ using Eq. (\ref{swF}) then implies that
\begin{align*}
{\mathcal I}
&=-D\int_{\Omega} \partial_{\sigma} G(\x,s+\alpha+\beta|\y) \times\int_{\partial \Omega} \partial_{\sigma'} G(\x',s|\y)f(\x',\z,s)d\x'd\y\\
&=-D\partial_\sigma \int_{\partial \Omega} \partial_{\sigma'} H(\x',s|\x,s+\alpha+\beta)f(\x',\z,s)d\x'\nonumber \\
&\equiv  \overline{\L}_{s,s+\alpha+\beta}[f](\x,\z,s),\ \x\in \partial \Omega,
\end{align*}
where
\begin{equation}
H(\x',s|\x,\tau)\equiv  \int_{\Omega}  G(\x',s|\y)G(\x,\tau|\y)d\y,
\end{equation}
and $\overline{L}_{s,\tau}$ is a second Dirichlet-to-Neumann operator on $\partial \Omega$.
We can thus write Eq. (\ref{woo}) in the more compact form
\begin{align}
\label{swcond1}
\alpha \overline{\L}_{s,s+\alpha+\beta}[f]+\L_{s+\alpha+\beta}[f_1] +z_1f_1=-\rho_1 \partial_{\sigma} G.
\end{align}

\noindent (iv) The second self-consistency condition is obtained by setting $\calF_0=\calF-\calF_1$ in Eq. (\ref{swhomT}c) with $j=0$:
\begin{align}
&\partial_{\sigma}\calF(\x,z,s)-\partial_{\sigma}\calF_1(\x,z,s)+z_0(\calF(\x,z,s)-\calF_1(\x,z,s))\nonumber \\
& \quad =-\rho_0 \partial_{\sigma} G(\x,s|\x_0) ,\   \x\in \partial \Omega.
\end{align}
This can be rearranged to give
\begin{align}
&\partial_{\sigma}\calF(\x,z,s)+z_0\calF(\x,z,s) \\
& \quad =(z_0-z_1) f_1(\x,z,s)-  \partial_{\sigma} G(\x,s|\x_0) ,\   \x\in \partial \Omega.\nonumber
\end{align}
Finally, using Eq. (\ref{swF}) and the definition (\ref{DtoN}) of the Dirichlet-to-Neumann operator $\L_s$, we have
\begin{equation}
\label{swcond2}
\L_s[f] +z_0f =(z_0-z_1)f_1-\partial_{\sigma}G.
\end{equation}

\noindent (v) In the 1D case the boundary $\partial \Omega$ is a single point so the resulting operator equations are simply scalars. Eqs. (\ref{swcond1}) and (\ref{swcond2}) thus reduce to Eqs. (\ref{cond1}) and (\ref{cond2}), respectively, and we can solve for $f$ and $f_1$ without the need for any spectral decompositions.
For $d>1$, we substitute the eigenvalue expansions (\ref{eig2}) and
\begin{equation}
\label{eig3}
f_1(\x,z,s)=\sum_{m=0}^{\infty}f_{1,m}(z,s) v_m(\x,s)
\end{equation}
into Eqs. (\ref{swcond1}) and (\ref{swcond2}) and take the inner product with the adjoint eigenfunction $v_n^*(\x,s)$. First, Eq. (\ref{swcond2}) reduces to the form
\begin{equation}
\label{sspec0}
(\lambda_n(s)+z_0)f_n(s) =(z_0-z_1)f_{1,n}(s)+\frac{1}{D}{\mathcal V}_n(s),
\end{equation}
with ${\mathcal V}_n$ defined in Eq. (\ref{calV}). Second Eq. (\ref{swcond1}) becomes
\begin{equation}
\label{sspec1}
[\lambda_n(s+\alpha+\beta)+z_1]f_{1,n}(s) +\sum_{m\geq 0} H_{nm}(s)f_m(s) =\frac{\rho_1}{D}{\mathcal V}_n(s),
\end{equation}
where
\begin{align}
 H_{nm}(s)&=-D\int_{\partial \Omega} v_n^*(\x,s)\partial_{\sigma}\left \{\int_{\partial \Omega}v_m(\x',s)\partial_{\sigma'}H(\x',s|\x,s+\alpha+\beta) d\x'\right \}d\x.
\label{H}
\end{align}
The orthogonality condition 
\begin{equation}
\int_{\partial \calM} v_n^*(\x,s)v_m(\x,s)d\x=\delta_{m,n}
\end{equation}
means that $v_n^*$ and $v_m$ can each be taken to have dimensions of [Length]$^{-(d-1)/2}$. It also follows that $H_{nm}(s)$ has dimensions of inverse length.
\medskip

\noindent (vi) Use Eq. (\ref{sspec0}) to express $f_n(s)$ in terms of $f_{1,n}(s)$ and substitute the result into Eq. (\ref{sspec1}):
\begin{align}
&[\lambda_n(s+\alpha+\beta)+z_1]f_{1,n}(s) +\sum_{m\geq 0} H_{nm}(s)\frac{(z_0-z_1)f_{1,m}(s)+ {\mathcal V}_m(s)/D}{ \lambda_m(s)+z_0}\nonumber \\
&\qquad =\frac{\rho_1}{D}{\mathcal V}_n(s).
\end{align}
Finally, introducing the vectors ${\bf f}_1(s)=(f_{1,n}(s), n\geq 0)$ and ${\bf g}(s)=({\mathcal V}_n(s)/D, n\geq 0)$, we can formally write the solution for ${\bf f}_1(s)$ as
\begin{align}
\label{eff}
{\bf f}_1(s)&=\left [{\bf M}(s+\alpha+\beta,z_1) +(z_0-z_1){\bf H}(s){\bf M}(s,z_0)^{-1} \right ]^{-1} \nonumber \\
& \quad \times \left [\rho_1{\bf I}-{\bf H}(s){\bf M}(s,z_0)^{-1}\right ]{\bf g}(s),
\end{align}
where ${\bf H}(s)$ is the matrix with elements $H_{nm}(s)$ and ${\bf M}(s,z)=\mbox{diag}(\lambda_1(s)+z,\lambda_2(s)+z,\ldots)$. Note that Eq. (\ref{eff}) is the higher-dimensional version of the solution (\ref{f1}).

\subsection{Partially absorbing sphere}

One example where the spectral decompositions of the Dirichlet-to-Neumann operator $\L_s$ is known exactly is a partially absorbing sphere. Let $\Omega=\{\x\in \R^3,\, 0 <  |\x| <R\}$ so that $\partial \Omega= \{\x\in \R^3,\,  |\x| =R\}$. The rotational symmetry of $\Omega$ implies that if $\L_s$ is expressed in spherical polar coordinates $(\rho,\theta,\phi)$, then the eigenfunctions are given by spherical harmonics, and are independent of the Laplace variable $s$ and the radius $\rho$:
\begin{equation}
v_{nm}(\theta,\phi)=\frac{1}{R} Y_n^m(\theta,\phi),\quad n\geq 0, \ |m|\leq n.
\end{equation}
From orthogonality, it follows that the adjoint eigenfunctions are
\begin{equation}
v^*_{nm}(\theta,\phi)=\overline{v}_{nm}^*(\theta,\phi)=(-1)^m\frac{1}{R} Y_n^{-m}(\theta,\phi).
\end{equation}
(Note that eigenfunctions are labeled by the pair of indices $(nm)$.)
The corresponding eigenvalues are \cite{Grebenkov19b}
\begin{equation}
\lambda_n(s)=\mu(s)\frac{i_n'(\mu(s) R)}{i_n(\mu(s) R)},
\end{equation}
where $\mu(s)=\sqrt{s/D}$ and $i_n$ is a spherical modified Bessel function of the first kind.
Since the $n$th eigenvalue is independent of $m$, it has a multiplicity $2n+1$.
It is also possible to compute the projection of the boundary flux in (\ref{calV}) by using appropriate series expansion of the corresponding Green's function. In particular, one finds that \cite{Grebenkov19b,Grebenkov20}
\begin{equation}
-D\partial_{\sigma}G(\x',s|\x)=\sum_{n=0}^{\infty} \frac{2n+1}{4\pi R^2} P_n(\x'\cdot \x/(\rho R))\frac{i_n(\mu(s) \rho)}{i_n(\mu(s) R)}
\end{equation}
with $|\x'|=R$, $|\x|=\rho <R$, and $P_n(x)$ a Legendre polynomial. Hence, since $\partial_{\sigma'}=-\partial/\partial \rho'$, we have
 \begin{align}
{\mathcal V}_{nm}(\x,s)&\equiv D\int_{|\x'|=R} v_{nm}(\theta',\phi')\frac{\partial}{\partial \rho'}G(\x',s|\rho,\theta,\phi)d\x'\nonumber \\
&=-v_{nm}(\theta,\phi)\frac{i_n(\mu(s) \rho)}{i_n(\mu(s) R)},
\end{align} 
with $\x=(\rho,\theta,\phi)$, $\x'=(\rho',\theta',\phi')$, and $\rho<R$. 
Finally, the matrix ${\bf H}(s)$ in equation (\ref{H}) becomes, after setting $\y=( \bar{\rho},\bar{\theta},\bar{\phi})$,
\begin{align}
&H_{nm,n'm'}(s)\nonumber \\
&=-D \int_{\Omega}d\y\,\bigg \{\int_{\partial \Omega} d\x\, v_{nm}^*(\theta,\phi) \int_{\partial \Omega}d\x'\, v_{n'm'}(\theta',\phi')\nonumber \\
&\qquad  \times \frac{\partial}{\partial \rho'}G(\x',s|\y) \frac{\partial}{\partial \rho}G(\x,s+\alpha+\beta|\y) \bigg \}\nonumber \\
&=-D\int_{\Omega}d\y\,\bigg \{v_{nm}(\bar{\theta},\bar{\phi})v_{n'm'}(\bar{\theta},\bar{\phi})\nonumber \\
&\hspace{2cm} \times\frac{i_n(\nu(s) |\y|)}{i_n(\nu(s) R)}\frac{i_{n'}(\mu(s) |\y|)}{i_{n'}(\mu(s) R)}\bigg \}
\nonumber \\
\label{H2}
&=-D\int_{0}^R\left [\frac{i_n(\nu(s)  \bar{\rho})}{i_n(\nu(s) R)}\frac{i_{n'}(\mu(s)  \bar{\rho})}{i_{n'}(\mu(s) R)}\right ] \bar{\rho}^2 d \bar{\rho} \\
&\hspace{2cm} \times \left [\int_{\partial \Omega} v_{nm}^*(\bar{\theta},\bar{\phi})v_{n'm'}(\bar{\theta},\bar{\phi})d\x\right ]\nonumber \\
&=-D\delta_{n,n'}\delta_{m,m'} \int_{0}^R\left [\frac{i_n(\nu(s) \bar{\rho})}{i_n(\nu(s) R)}\frac{i_n(\mu(s)  \bar{\rho})}{i_n(\mu(s) R)}\right ] \bar{\rho}^2 d \bar{\rho}.\nonumber
\end{align}
That is, ${\bf H}$ is a diagonal matrix.

 We conclude that in the case of a sphere, one can obtain explicit expressions for the doubly Laplace-transformed propagators. However, in order to incorporate a non-exponential local time distribution $\Psi_j(\ell_j)$ for partial absorption in the state $N_t=j$, it is necessary to invert the Laplace transform with respect to $\z$. In general, this would have to be implemented numerically.

\setcounter{equation}{0}\section{Discussion} In this paper we combined two distinct sources of complexity in diffusion-mediated surface absorption. The first involves a general probabilistic rule for partial absorption, in which the surface reactivity depends on the number of encounters between a diffusing particle and the surface. The encounter rate is determined by the boundary local time, and partial absorption can be formulated mathematically in terms of the generalized propagator $P(\x,\ell,t)$. The second type of complexity arises when there is random switching between two distinct surface boundary conditions, which we assume is due to the particle switching between two different conformational states labeled by $j=0,1$. If at least one of the boundary conditions is partially absorbing, then it is necessary to introduce a pair of boundary local times $\ellb=(\ell_0,\ell_1)$, which keep track of surface-particle encounters in each of the discrete states, and a corresponding pair of generalized propagators $P_j(\x,\ellb,t)$, $j=0,1$. The latter evolve according to a system of differential CK equations that can be solved by performing a double Laplace transform with respect to $\ell_0$ and $\ell_1$. 

One major assumption of the hybrid model was that the surface-particle interactions in the different discrete states were statistically uncorrelated. This allowed us to define a probabilistic rule for partial absorption in which the stopping local time distribution decomposed into the product $\Psi(\ell_0,\ell_1)=\Psi_0(\ell_0)\Psi_1(\ell_1)$. That is, the probability of absorption when the particle was in the discrete conformational state $j\in \{0,1\}$ only depended on the local time accumulated whilst in that state. Such a rule reduced to switching Robin boundary conditions in the case of constant reactivities. In future work it would be interesting to explore probabilistic rules for which $\Psi(\ell_0,\ell_1)\neq \Psi_0(\ell_0)\Psi_1(\ell_1)$. However, the physical interpretation of the resulting switching absorption process is less clear. 

Another possible extension of the theory would be to treat the interior of the bounded domain $\Omega$ in Fig. \ref{fig2} as a partially absorbing substrate or trap. The diffusing particle can now freely enter and exit $\Omega$, and the probability of being absorbed depends on the amount of time spent within $\Omega$ (in the absence of switching). The latter is specified by another Brownian functional known as the occupation time $A_t$ \cite{Majumdar05}. We have recently shown how to extend the encounter-based approach to partially absorbing substrates without switching by constructing the generalized propagator for the occupation time $A_t$ rather than the local accumulation time $\ell_t$ \cite{Bressloff22a}. Moreover, the corresponding propagator BVP can be solved by computing the spectral decomposition of an associated Dirichlet-to-Neumann operator \cite{Bressloff22b}. Following along analogous line to partially absorbing surfaces, we could take into account stochastic switching between different conformational states by introducing a corresponding set of state-dependent occupation times and generalized propagators. The latter would evolve according to a system of differential CK equations that is the analog of Eqs. (\ref{swPloc}).

Finally, a number of recent statistical analyzes of single-particle tracking (SPT) experiments \cite{Das09,Persson13,Slator15} suggest that proteins within living cells can switch between different discrete states with different diffusivities. Such switching could be due to interactions between proteins and the actin cytoskeleton \cite{Das09} or due to protein-lipid interactions \cite{Yamamoto17}. These observations have motivated several analytical studies of Brownian particles with switching diffusivities \cite{Akimoto16,Godec17,Bressloff17a,Bressloff19a}. It is typically assumed that when a particle is in the discrete conformational state $N_t =j\in \{0,1\}$, its corresponding diffusivity is $D_j$ with $D_0\neq D_1$. One can then introduce a corresponding pair of probability densities $p_j(\x,t)$, $j=0,1$, which evolve according to a differential CK equation that takes into account transitions between the discrete states. However, such models do not incorporate the effects of particle-surface interactions that may play a role in such switching. A modified version of our switching propagator model could be one way to take into account such interactions.


\bigskip

\begin{thebibliography}{}




\bibitem{Metzler14} R. Metzler, G. Oshanin, and S. Redner (Eds.) First-
Passage Phenomena and Their Applications (Singapore:
World Scientific, 2014).

\bibitem{Benichou14} O. Benichou and R. Voituriez, From first-passage times
of random walks in confinement to geometry-controlled
kinetics, Phys. Rep. {\bf 539} 225-284 (2014)

\bibitem{Bressloff13} P. C. Bressloff and J. M. Newby, Stochastic models of
intracellular transport, Rev. Mod. Phys. {\bf 85}, 135-196
(2013).

\bibitem{Grebenkov19a} {D. S. Grebenkov} {Imperfect Diffusion-Controlled Reactions.}
in {\em Chemical Kinetics: Beyond the Textbook} Eds.
Lindenberg K, Metzler R and Oshanin G World Scientific (2019).



\bibitem{Bressloff22} P. C. Bressloff, {\em Stochastic Processes in Cell Biology} (2nd. edition) Springer, Switzerland (2022).

\bibitem{Note1} In physical chemistry there is an important distinction between {\em absorption} and {\em adsorption}. The former is defined to be a bulk process in which a fluid is dissolved into a liquid or solid (absorpent). On the other hand, adsorption is a surface phenomenon in which atoms, ions or molecules adhere to the surface of the adsorpent. Therefore, technically speaking, we should refer to reactive surfaces as adsorpents rather than absorpents. However, since {\em absorption} is the more common term used to describe boundary conditions in the mathematical analysis of diffusion processes, we use this terminology throughout.

 
\bibitem{Levy39} {P. L\`evy} {\em Sur certaines processus stochastiques homogenes.} {Compos. Math.} {\bf 7} (1939) 283.



 
 \bibitem{McKean75} {H. P. McKean} {\em Brownian local time.} {Adv. Math.} {15} (1975) 91-111.
 
  
  \bibitem{Majumdar05} {S. N. Majumdar} {\em Brownian functionals in physics and computer science.} {Curr. Sci.} {89} (2005) 2076.
  
  \bibitem{Freidlin85} {M. Freidlin} {Functional Integration and Partial Differential Equations}
Annals of Mathematics Studies. Princeton University Press, Princeton
New Jersey (1985)


 \bibitem{Milshtein95} {G. N. Milshtein} {\em The solving of boundary value problems by numerical
integration of stochastic equations.} {Math. Comp. Sim.} {38} (1995) 77-85.

 \bibitem{Papanicolaou90} {V. G. Papanicolaou} {\em The probabilistic solution of the third boundary
value problem for second order elliptic equations} {Probab. Th. Rel. Fields}
{87} (1990) 27-77.


\bibitem{Singer08} {A. Singer, Z. Schuss, A. Osipov and D. Holcman.} {\em Partially reflected
diffusion.} {SIAM J. Appl. Math.} {68} (2008) 844-868.

\bibitem{Bartholomew01} C. H. Bartholomew, Mechanisms of catalyst deactivation,
Appl. Catal. A: Gen. {\bf 212}, 17-60 (2001).


\bibitem{Filoche08} M. Filoche, D. S. Grebenkov, J. S. Andrade Jr., and B.
Sapoval, Passivation of Irregular Surfaces Accessed by
Diffusion, Proc. Natl. Acad. Sci. {\bf 105}, 7636-7640 (2008).

\bibitem{Grebenkov20} {D. S. Grebenkov}  {Paradigm shift in diffusion-mediated surface phenomena.} {Phys. Rev. Lett.} {\bf 125} 078102 (2020).



\bibitem{Grebenkov22} {D. S. Grebenkov}  {An encounter-based approach for restricted diffusion with a gradient drift.}  J. Phys. A {\bf 55} 045203 (2022) .

\bibitem{Bressloff22a}  {P. C. Bressloff}. {Diffusion-mediated absorption by partially reactive targets: Brownian functionals and generalized propagators.} J. Phys. A. {55} (2022) 205001.
 

\bibitem{Reingruber10} {J. Reingruber and D. Holman} Narrow escape for a stochastically gated Brownian ligand. J. Phys. Cond. Matter {\bf 22} 065103 (2010)

\bibitem{Bressloff15a} P. C. Bressloff and S. D. Lawley,
Moment equations for a piecewise deterministic PDE 
{J. Phys. A } {\bf 48} 10500 (2015).

\bibitem{Bressloff15c} P. C. Bressloff and S. D. Lawley,
Escape from subcellular domains with randomly switching boundaries. 
Multiscale Model. Simul. {\bf 13} 1420-1445 (2015). 

\bibitem{Evans20} M. R. Evans, S. N. Majumdar, and G. Schehr, Stochastic resetting and applications. J. Phys. A {\bf 53} 193001 (2020).



\bibitem{Evans11a} M. R. Evans and S. N. Majumdar, Diffusion with stochastic resetting, Phys. Rev. Lett. {\bf 106} 160601 (2011).

\bibitem{Evans11b} M. R. Evans and S. N. Majumdar, Diffusion with optimal resetting, J. Phys. A Math. Theor. {\bf 44} 435001 (2011).


\bibitem{Evans14} M. R. Evans and S. N. Majumdar, Diffusion with resetting
in arbitrary spatial dimension, J. Phys. A {\bf 47}, 285001 (2014).

\bibitem{Bressloff20} P. C. Bressloff. Diffusive search for a stochastically-gated target with resetting. J. Phys. A {\bf 53} 425001 (2020)


\bibitem{Boyer21} G. Mercado-Vasquez and D. Boyer, Search of stochastically gated targets with diffusive particles under resetting. J. Phys. A: Math. Theor. {\bf 54} 444002 (2021)

\bibitem{Bressloff22c} P. C. Bressloff, Diffusion-mediated surface reactions and stochastic resetting. {J. Phys. A.} In press. arXiv:2202.01119 (2022)

\bibitem{note2} Local time resetting can be implemented by associating with each discrete conformational state $n$ an additional internal state variable $\sigma_n$ that increases monotonically with the corresponding local time $\ell_n$. If the probability of absorption depends on $\sigma_n$, then local time resetting is mathematically equivalent to resetting the internal states. This idea was previously introduced within the context of partially absorbing boundaries without switching \cite{Bressloff22c}.


 \bibitem{Newby10}
J. M. Newby and P. C. Bressloff, Quasi-steady state reduction of molecular-based models of directed intermittent search.
\newblock Bull Math Biol \textbf{72} 1840-1866 (2010).



\bibitem{Zmurchok17} C. Zmurchok, T. Small, M. Ward and L. Edelstein-Keshet,  Application of quasi-steady state methods to nonlinear models of intracellular transport by molecular motors. Bull. Math. Biol. {\bf 79} 1923-1978 (2017).

\bibitem{Bressloff19a} P. C. Bressloff, S. D. Lawley and P. Murphy, Protein concentration gradients and switching diffusions. Phys. Rev. E {\bf 99} 032409 (2019).

\bibitem{Lawley15} S. D. Lawley and J. P. Keener, A new derivation of Robin boundary conditions through homogenization of a stochastically switching boundary. SIAM J. Appl. Dyn. Syst. {\bf 14} 1845-1867 (2015).

\bibitem{note3} Interestingly, the dependence of the solution (\ref{2calPsol}) on the weighted mean $\overline{z}$ provides an alternative way of understanding how to derive Robin boundary conditions via the homogenization of a stochastically switching boundary. As shown in Ref. \cite{Lawley15}, when a particle diffuses in a domain with a boundary that switches between Dirichlet (for $n=0$) and Neumann (for $n=1$), the fast switching limit typically yields a Dirichlet boundary condition rather than a mixed or Robin boundary condition. One exception is when $\rho_1 \rightarrow 1$ as $\alpha,\beta \rightarrow \infty$. In terms of $\overline{z}=\rho_0 z_0+\rho_1 z_1$, we see that if $\rho_0=1-\rho_1 >0$ then $\overline{z}\rightarrow \infty$ as $z_0\rightarrow \infty$ (both of which correspond to a Dirichlet boundary condition). On the other hand, $\overline{z}$ can remain finite in the fast switching limit if $\rho_0\rightarrow 0$ and $\rho_1\rightarrow 1$ in an appropriate way.

\bibitem{Grebenkov19b} D. S. Grebenkov, {Spectral theory of imperfect diffusion-controlled
reactions on heterogeneous catalytic surfaces}
{J. Chem. Phys.} {\bf 151} 104108 (2019)

\bibitem{Bressloff22b} P. C. Bressloff,  Spectral theory of diffusion in partially absorbing media. arXiv:2205.08929  (2022)

\bibitem{Das09} R. Das, C. W. Cairo and D. A. Coombs, A hidden Markov model for single particle tracks quantifies dynamic interactions between LFA-1 and the actin cytoskeleton. 
PLoS Comp. Biol. {\bf 5} e1000556 (2009).

 \bibitem{Persson13} F. Persson, M. Linden, C. Unoson and J. Elf, Extracting intracellular diffusive states and transition rates from single-molecule tracking data. Nat. Meth. {\bf 10} 265-269 (2013).
 
 \bibitem{Slator15}
P. J. Slator, C. W. Cairo, and N. J. Burroughs, Detection of diffusion heterogeneity in single particle tracking trajectories using a hidden Markov model with measurement noise propagation. PLoS ONE {\bf 10} e0140759 (2015). 


  \bibitem{Yamamoto17}  E. Yamamoto, T. Akimoto, A. C. Kalli, K. Yasuoka and M. S. P. Sansom, M. S. P.: Dynamic interactions between a membrane binding protein and lipids induce fluctuating diffusivity. Sci. Adv. {\bf 3} e1601871 (2017).
  
  \bibitem{Akimoto16} T. Akimoto and E. Yamamoto, Distributional behaviors of time-averaged observables in the Langevin equation with fluctuating diifusivity: normal diffusion but anomalous fluctuations. Phys. Rev. E {\bf 93} 062109 (2016).

\bibitem{Godec17} A. Godec and R. Metzler, First passage time statistics for two-channel diffusion. J. Phys. A {\bf 50} 084001 (2017). 

 \bibitem{Bressloff17a} P. C. Bressloff and S. D. Lawley, Temporal disorder as a mechanism for spatially heterogeneous diffusion
Phys. Rev. E {\bf 95} 060101(R) (2017).



\end{thebibliography}
\end{document}